\documentclass[aps,prc,twocolumn,longbibliography,floatfix,10pt]{revtex4}
\usepackage{graphicx}
\usepackage{dcolumn}
\usepackage{bm}
\usepackage{amssymb}
\usepackage{amsmath,bm}
\usepackage[normalem]{ulem}
\usepackage{multirow}

\usepackage[usenames,dvipsnames]{xcolor}
\usepackage{lipsum}

\renewcommand{\sout}{\bgroup \color{red} \ULdepth=-.5ex \ULset}

\begin{document}


\title{{Charged pion production from Au + Au collisions at $\sqrt{s_{NN}}=2.4$ GeV in the Relativistic Vlasov-Uehling-Uhlenbeck model}}

\author{Kyle Godbey\footnote%
{godbeyky@msu.edu}}    
\affiliation{Cyclotron Institute, Texas A$\&$M University, College Station, TX 77843, USA}
\altaffiliation{Current address: Facility for Rare Isotope Beams, Michigan State University, East Lansing, Michigan 48824, USA }

\author{Zhen Zhang\footnote%
{zhangzh275@mail.sysu.edu.cn}}
\affiliation{Sino-French Institute of Nuclear Engineering and Technology, Sun Yat-sen University, Zhuhai 519082, China}

\author{Jeremy W. Holt\footnote%
{holt@comp.tamu.edu}}
\affiliation{Cyclotron Institute and Department of Physics and Astronomy, Texas A$\&$M University, College Station, TX 77843, USA}

\author{Che Ming Ko\footnote%
{ko@comp.tamu.edu}}
\affiliation{Cyclotron Institute and Department of Physics and Astronomy, Texas A$\&$M University, College Station, TX 77843, USA}

\date{\today}

\begin{abstract}
Using the isospin-dependent relativistic Vlasov-Uehling-Uhlenbeck (RVUU) model, we study charged pion ($\pi^\pm$) production in Au+Au collisions at $\sqrt{s_{NN}}=$ 2.4 GeV.  By fitting the density dependence of the $\Delta$ resonance production cross section in nuclear medium to reproduce the experimental $\pi^\pm$ multiplicities measured by the HADES Collaboration, we obtain a good description of the rapidity distributions and transverse momentum spectra of $\pi^\pm$ in collisions at various centralities. Some shortcomings in the description of $\pi^{\pm}$ production may indicate the need for including the strong potential on $\pi^\pm$ in RVUU, which is at present absent. We also calculate the proton rapidity distribution in the most central collisions and compare with the coalescence invariant proton rapidity distribution extracted from preliminary HADES data.  
\end{abstract}

\maketitle

\section{Introduction}

The study of pion production in heavy-ion collisions has a long and storied history as it is the lightest hadron produced in these reactions. From studies based on the Boltzmann-Uehling-Uhlenbeck (BUU) transport model~\cite{Bertsch:1988ik}, it was found that the inclusion of an equation of state (EOS) through the mean-field potentials acting on nucleons was essential to describe the pion yield produced in heavy-ion collisions~\cite{Kruse:1985hy}. Unfortunately, the difference between the pion yields from the use of a soft and a stiff EOS is not large, making it difficult to extract from experimental pion data the information on the stiffness of the nuclear EOS.  Also, very different results on the pion yield as well as the pion rapidity distribution and transverse momentum spectrum are predicted from various transport models based on the BUU or the quantum molecular dynamics (QMD) approach~\cite{Kolomeitsev:2004np}. 

More recently, the study of the $\pi^-/\pi^+$ yield ratio in heavy ion collisions at energies near the pion production threshold has drawn great 
attentions~\cite{Xiao2009,Feng2010,Xie2013,Xu2010,Xu2013,Cozma2017,Zhang2017,Li2015,Ferini2005,Song2014,Yong:2017cdl,Ikeno2016} as it was suggested in Ref.~\cite{li2002} that the stiffness of the nuclear symmetry energy $E_{\rm sym}(\rho)$ at high densities could have appreciable effect on this ratio. In this case, studying pion production in heavy ion collisions can greatly impact the progress of nuclear physics as the nuclear symmetry energy is at the heart of a complete understanding of nuclear physics on various scales -- governing the detail and structure of static nuclei~\cite{vinas2014,baldo2016,Adhikari:2021phr}, transfer and equilibration in low-energy nuclear reactions~\cite{baran2005,li2008,godbey2017,simenel2020}, and the structure and dynamics of neutron stars~\cite{lattimer2004,steiner2005,lattimer2007,li2019}. Despite this outsized impact on a wide range of physical phenomena, the high-density behavior of $E_{\rm sym}(\rho)$ has not been as well constrained as other properties of nuclear matter  (see \cite{li2008} for a comprehensive review).   

In heavy-ion collisions at beam energy $E/A$ below 1.3 GeV, pions are dominantly produced from the decay of $\Delta$ resonances formed from $NN$ inelastic scattering in the high density regions during the collision. As the $nn\rightarrow p\Delta^-$ and $pp\rightarrow n\Delta^{++}$ channels dominate the final charged pion multiplicities, the ratio of charged pions can provide insight into the isospin asymmetry of the dense matter in which they are formed. In practice, however, studies using various transport models were able to describe the pion data measured by the FOPI Collaboration~\cite{Reisdorf:2006ie} with nuclear symmetry energy at high densities ranging from super soft~\cite{Xiao2009,Xie2013} to stiff ones~\cite{Feng2010,Cozma2017}. This has led to the transport model evaluation project (TMEP) to understand the origin for the different predictions from a large number of transport models~\cite{Xu:2016lue,Zhang:2017esm,Ono:2019ndq}. Some of these transport models have also recently been used to predict the charged pion multiplicities for reactions between neutron-rich Sn isotopes at a beam energy of 270 AMeV, though none of them can describe the available data from the S$\pi$RIT Collaboration~\cite{Jhang:2020rps}. Although a subsequent study of the S$\pi$RIT pion data based on the dcQMD model~\cite{Cozma:2021tfu} has put a constraint on the density slope  of the nuclear symmetry energy at normal saturation density~\cite{Estee:2021khi}, the large uncertainty in the extracted value indicates the need for a better understanding and modeling of pion production in nuclear collisions.

Recently, charged pions produced from Au+Au collisions at a beam energy of 1.23~AGeV or center of mass energy $\sqrt{s_{NN}}=2.4$~GeV have  been measured by the HADES Collaboration~\cite{hades2020}.  Although the collision energy in this reaction is too high for the nuclear symmetry energy to affect the charged pion ratio, it adds to the available experimental data to benchmark model calculations and improve theoretical descriptions of pion production in heavy-ion collisions. Indeed, predicted pion yields from the five transport models used in studying the HADES pion data not only fail to describe the data but also disagree among themselves~\cite{Hartnack:1997ez,Cassing:1999es,Aichelin:2019tnk,Buss:2011mx,Petersen:2018jag}. However, the yield of neutral pions from the GiBUU model in these five transport models has been found to give the correct contribution to the low mass dilepton spectrum measured by the HADES Collaboration~\cite{HADES:2019auv} from their Dalitz decays~\cite{Larionov:2020fnu}.  Despite this tension between the pion and dilepton data from the HADES Collaboration, to find the possible reasons for the failure of these transport models to describe the HADES pion data is still of great interest. The  current work builds upon recent extensions to the relativistic Vlasov-Uehling-Uhlenbeck (RVUU) model~\cite{Song2015,Zhang2017} in an attempt to better describe $\pi^\pm$ production in this reaction.

The structure of the paper is as follows:  Sec.~\ref{sec:methods} exhibits an overview of the RVUU model and how it is used to determine pion production, along with the modifications to the in-medium density dependence of the $N+N$ inelastic cross section introduced in the current work. Sec.~\ref{sec:results} then presents the results from RVUU simulations in comparison with experimental data from Ref.~\cite{hades2020}. Finally, Sec.~\ref{sec:conclusions} summarizes the results and discusses future directions.

\section{Relativistic Vlasov-Uehling-Uhlenbeck Model}
\label{sec:methods}

In the present study, we use the relativistic Vlasov-Uehling-Uhlenbeck (RVUU) model  to further probe in-medium effects in heavy-ion collisions. The RVUU model describes the time evolution of the  nucleonic phase-space distribution function, $f({{\bf r}, {\bf p}, t})$, by using the following transport equation,
\begin{equation} \label{eq:rvuu}
  \partial_t f + {\bf v} \cdot {\boldsymbol\nabla} r f -{\boldsymbol\nabla}_r H \cdot {\boldsymbol\nabla}_p f = I_c.
 \end{equation}
In the above, $H$ is the mean-field Hamiltonian of a nucleon, which was originally based on relativistic energy functionals including only isoscalar scalar $\sigma$ and vector $\omega$ meson fields ~\cite{ko1987,ko1988}, and has been later  extended to also include the isovector scalar $\delta$ and vector $\rho$ fields~\cite{Song2015}. The resulting mean-field Hamiltonian~\cite{Song2015} in Eq.~(\ref{eq:rvuu}) is then given by
\begin{equation}\label{hamiltonian}
H= \sqrt{m_N^{*2} + {\bf p}^{*2}} + g_\omega \omega^0 \pm g_\rho\rho_3^0,
\end{equation}
for protons ($+$) and neutrons ($-$) of in-medium mass $m_N^*=m_N-(g_\sigma\sigma\pm g_\delta\delta_3)$ and kinetic momentum ${\bf p}^*={\bf p}+g_\omega{\boldsymbol\omega}\pm g_\rho{\boldsymbol\rho_3}$ in the center-of mass frame of the colliding nuclei.  The meson fields $\sigma$, $\delta_3$, $\omega^\mu$ and $\rho_3^\mu$ in the above equations are related to the proton and neutron scalar densities and their currents.  Assuming the same  interactions of the $\Delta$ resonance, whose isospin is $T=3/2$, with the $\sigma$ and $\omega$ fields as those for nucleons and relating its interactions with the $\delta$ and $\rho$ fields to those of nucleons via its isospin structure, the RVUU model has also been used to describe the evolution of the $\Delta$ resonances in heavy ion collisions. For the values of the coupling constants $g_\sigma$, $g_\omega$, $g_\delta$ and $g_\rho$ as well as the nonlinear coupling constants of the scalar fields, we take the NL$\rho$ parameter set given in Ref.~\cite{Liu2002}, which can well reproduce our empirical knowledge on asymmetric nuclear matter.

The collision integral, $I_c$, in Eq.~(\ref{eq:rvuu}) describes the effect of nucleon-nucleon elastic and inelastic scatterings on the momenta of scattering nucleons and their probabilities to change to the $\Delta$ resonance. Governing the collision integral is thus the baryon-baryon elastic and inelastic cross sections.  For $pp$ and $np$ elastic scatterings,  RVUU uses the following parameterizations as in the Giessen Boltzmann–Uehling–Uhlenbeck model~\cite{Buss:2011mx},
\begin{widetext}
\begin{equation}
 \sigma_{\text {el }}^{p p}= \begin{cases}5.12 m_{N} /\left(s-4 m_{N}^{2}\right)+1.67 & \text { for } p_{\text {lab }}<0.435 \\ 23.5+1000\left(p_{\text {lab }}-0.7\right)^{4} & \text { for } 0.435<p_{\text {lab }}<0.8 \\ 1250 /\left(p_{\text {lab }}+50\right)-4\left(p_{\text {lab }}-1.3\right)^{2} & \text { for } 0.8<p_{\text {lab }}<2 \\ 77 /\left(p_{\text {lab }}+1.5\right) & \text { for } 2<p_{\text {lab }}<6\end{cases}
\end{equation}
and 
\begin{equation}
\sigma_{\text {el }}^{n p}= \begin{cases}17.05 m_{N} /\left(s-4 m_{N}^{2}\right)-6.83 & \text { for } p_{\text {lab }}<0.525 \\ 33+196\left|p_{\text {lab }}-0.95\right|^{2.5} & \text { for } 0.525<p_{\text {lab }}<0.8 \\ 31 / \sqrt{p_{\text {lab }}} & \text { for } 0.8<p_{\text {lab }}<2 \\ 77 /\left(p_{\text {lab }}+1.5\right) & \text { for } 2<p_{\text {lab }}<6,\end{cases}
\end{equation}
\end{widetext}
where $p_{\rm lab}$ in units of GeV$/c$ is the beam momentum in the laboratory frame, $m_N=0.939$ GeV is the bare nucleon mass, and the invariant energy $\sqrt{s}=\sqrt{2m_N^2+2m_N\sqrt{m_N^2+p^2_{\rm{lab}}}}$. The $nn$ elastic scattering cross section is taken to be the same as that for $pp$ scattering, and the $N\Delta$ elastic scattering cross sections are taken to be the average of those for $pp$ and $np$ scatterings. Taking into account the modifications of nucleon and $\Delta$ resonance masses in medium, the total elastic cross sections are calculated using the ``free" invariant energy $\sqrt{s_{\rm{free}}}=\sqrt{m_1^*+\bm{p}_1^*}+\sqrt{m_2^*+\bm{p}_2^*} -(m_1^*-m_1)-(m_2^*-m_2)$ as in Ref.~\cite{Buss:2011mx}, where $m_1$ ($m_1^*$) and $m_2$ ($m_2^*$) are the (effective) masses of the two scattering particles, and $\bm{p}_1^*$ and $\bm{p}_2^*$ are their respective kinetic momenta in their center of mass frame. We further take the differential cross sections in this frame to be isotropic.

For $\Delta$ resonance production in nucleon-nucleon scattering, the RVUU uses the total and differential cross sections calculated in the one-boson exchange model and parametrized as a function of $d (\rm{GeV}) = \sqrt{s}-\sqrt{s_{\rm{th}}}$ for the reactions $p+p\to n+\Delta^{++}$ and $n+n\to p+\Delta^-$~\cite{huber1994}, i.e.,
\begin{eqnarray}
&&\sigma_{NN\rightarrow N\Delta} (\rm{mb})\notag\\
&&= \left\{
\begin{array}{ll}
    1341d^{2.819}, & 0\leqslant d  \leqslant 0.186,\\
    18.51-235.2(d-0.356)^2, & 0.186 < d \leqslant 0.436,  \\
    1581(d+2.014)^{-4.957}, & 0.436 < d  \leqslant 2.486,\notag\\
\end{array}
\right.
\end{eqnarray}
and 
\begin{equation}
\frac{d\sigma_{NN\rightarrow N\Delta}}{d\cos\theta} \propto 
    \left\{
    \begin{array}{ll}
    \rm{exp}(b\rm{cos}\theta),~~&\text{forward}, \\
    \rm{exp}(-b\rm{cos}\theta),~~&\text{backward},\\
    \end{array}
    \right.
\end{equation}
with
\begin{widetext}
$$
b = \left\{ 
 \begin{array}{ll}
     19.71d^{1.551}, & 0\leqslant d \leqslant 0.416, \\
     33.41\rm{arctan}[0.5404(d-0.132)^{0.9784}],
     & 0.416< d \leqslant 2.486.
 \end{array}   
\right. 
$$
\end{widetext}
A factor of 1/3 is, however, multiplied to above $\sigma_{NN\to N\Delta}$ for the reactions $p+p\rightarrow p+\Delta^+$, $n+n\rightarrow n+\Delta^0$, $p+n\to p+\Delta^0$, and $p+n\to n+\Delta^+$ to take into account the isospin dependence of the $NN\to N\Delta$ reaction.

The threshold effects on $\Delta$ production cross sections are taken into account as in Ref.~\cite{Song2015}. For the mass of the produced $\Delta$
resonance, it is determined according to 
\begin{eqnarray}
P\left(m^{*}\right)=\frac{\mathcal{A}(m^*) p^{*}}{\int_{m^*_{\min }}^{m^*_{\max }} d m^{*\prime} \mathcal{A}\left(m^{*\prime}\right) p^{*}\left(m^{*\prime}\right)},
\end{eqnarray}
where $m^{*}$ is the effective mass of $\Delta$, and $m^*_{\min }$ and $m^*_{\text {max }}$ are the minimum and maximum effective masses of $\Delta$ that are allowed to form. The  $\mathcal{A}(m^*)$ is the in-medium spectral function  given by
\begin{eqnarray}
\mathcal{A}(m^*)=\frac{1}{\mathcal{N}} \frac{4 m_{0}^{*2} \Gamma}{\left(m^{*2}-m_{0}^{*2}\right)^{2}+m_{0}^{*2} \Gamma^{2}},
\end{eqnarray}
with $\mathcal{N}$ being the normalization factor and $m^*_0$ being the $\Delta$ pole mass of $1.232$ GeV shifted by the scalar-isoscalar potential. The total decay width of a $\Delta$ resonance of effective mass $m^*$ and in isospin state $m_T$ in its rest frame is taken to be
\begin{eqnarray}
\Gamma=\sum_{m_{t}} \frac{0.47 C q^{3}}{m_{\pi}^{2}+0.6 q^{2}},
\end{eqnarray}
where $m_{t}$ is the isospin state of the emitted pion, and $C=$ $\left|\left\langle\frac{3}{2}, m_{T} \mid 1, m_{t}, \frac{1}{2}, m_{T}-m_{t}\right\rangle\right|^{2}$ is the square of the Clebsch-Gordan coefficient from the isospin coupling. The magnitude of the momentum of the pion or nucleon in the $\Delta$ rest frame is denoted by $q$, and it is given by 
\begin{eqnarray}
q=\frac{\sqrt{\left[m{^{*2}}-\left(m_{N}^{*}+m_{\pi}\right)^{2}\right]\left[m^{*2}-\left(m_{N}^{*}-m_{\pi}\right)^{2}\right]}}{2 m^{*}}.
\end{eqnarray}
Note that in the calculation of the $\Delta$ decay width and spectral function, we only include the effect of the scalar potential and neglect the vector potential for simplicity.  For the reciprocal channel, $N+\Delta \rightarrow N+N$, its cross section is obtained by employing the detailed balance relation discussed in detail in Ref.~\cite{Zhang2017}.

\begin{figure}[h]
\includegraphics[width=8cm]{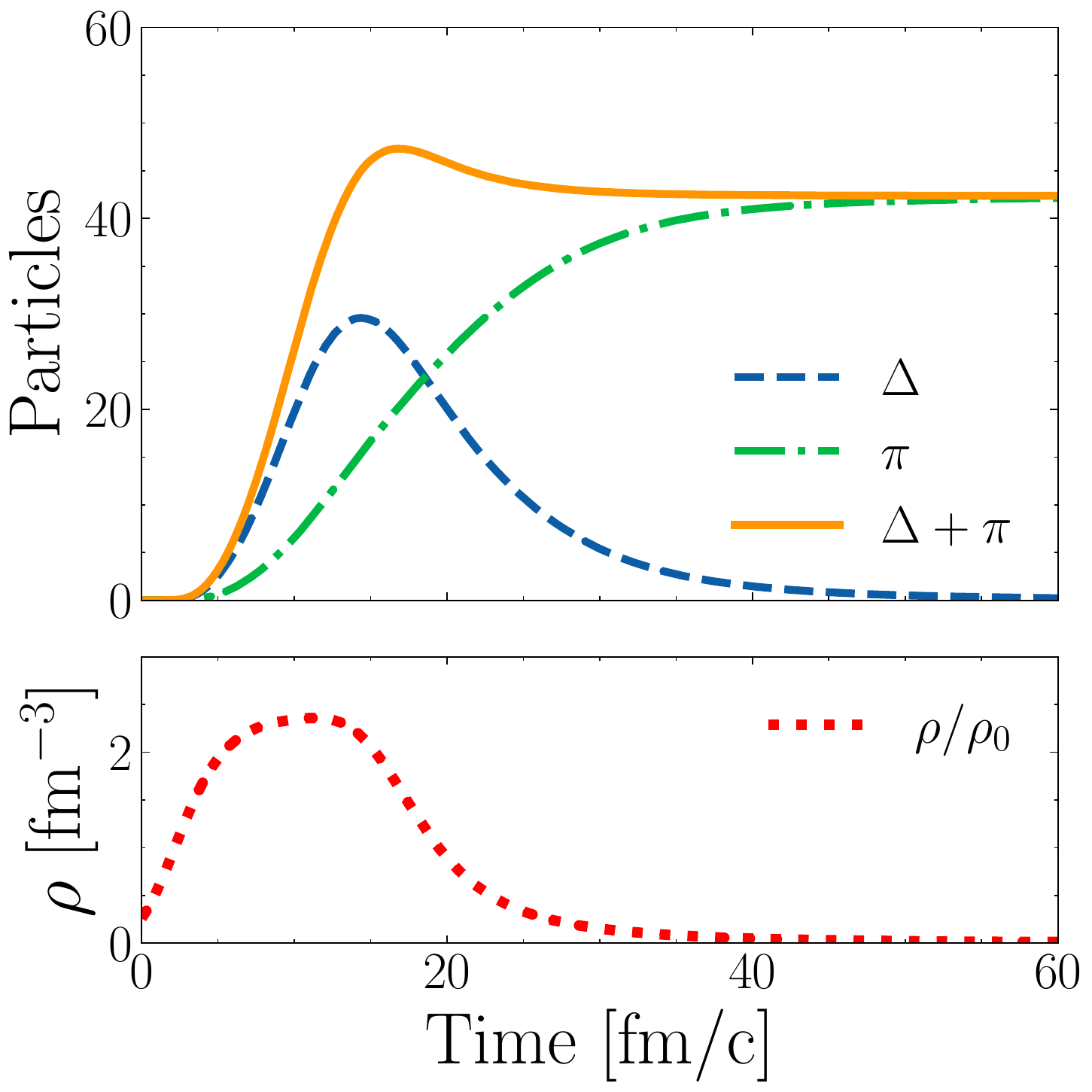}
\caption{\label{fig:pidelta}\protect(Color Online). Top window: number of pions (dot-dashed lines) and  Deltas (dashed lines) produced during RVUU simulations for the most central ($0\%-10\%$) Au+Au collisions at $\sqrt{s_{NN}}=2.4$ GeV over time. The solid line corresponds to the sum of Deltas and pions. Bottom window: maximal density in the central cells of the system at time $t$.}
\end{figure}

The final mechanism to consider is the decay of $\Delta$ resonances into $N\pi$ pairs and its inverse, $N+\pi\rightarrow\Delta$. The pions produced by $\Delta$ decay are propagated in time as free particles unless they scatter with nucleons to again form  $\Delta$ resonances. The $N$-$\pi$ inelastic scattering cross section is related to the decay width of the formed $\Delta$ resonance via the detailed balance relation~\cite{Zhang2017}. This naturally leads to a balance of $\Delta$ resonances and pions during the course of the reaction, with the sum of the two species being essentially constant after the period of highest density has passed. Figure~\ref{fig:pidelta} shows this balance as a function of time, representing the sum of $\pi^-,\pi^0,\textrm{ and }\pi^+$ particles as $\pi$ and the sum of $\Delta^-, \Delta^0, \Delta^+,\textrm{ and } \Delta^{++}$ resonances as $\Delta$.

With this description alone a reasonable result is obtained for both the dynamics and total particle production in nuclear reactions. However, as shown in previous studies with the RVUU model~\cite{Song2015, Zhang2017}, a reduction factor to the $\sigma_{NN\rightarrow N\Delta}$ cross sections, which approximates the  medium effects on $\Delta$ resonance production~\cite{engel1994,larionov2001,larionov2000}, is needed to describe pion production  in intermediate energy  heavy-ion collisions. The suppression of $NN$ inelastic scattering cross sections in nuclear medium is partly due to the modified phase-space in the final state and incident flux  in the initial state because of decreasing nucleon and $\Delta$ resonance masses (i.e., effective mass)  in the nuclear medium~\cite{Larionov:2003av}. In the relativistic mean-field model as used in RVUU, the inclusion of isovector-scalar $\delta$ meson leads to $m_p^*>m_n^*$ and $m_{\Delta^{++}}^*>m_{\Delta^{+}}^*>m_{\Delta^{0}}^*>m_{\Delta^{-}}^*$ in neutron-rich matter. Also, recent theoretical studies (see e.g., Refs.~\cite{LI2017557,Cui2021}) indicate that the exchange of the $\delta$ meson in the $NN\to N\Delta$ scattering can also cause a splitting of the suppression factors for the $\Delta$ production cross sections in different channels. To take into account these isospin-dependent suppression factors for the in-medium $\Delta$  production cross sections, we consider in the current work a modified scheme wherein the $\Delta^+$ and $\Delta^{++}$ channels are allowed to have a slightly different density dependence than that of the $\Delta^0$ and $\Delta^{-}$ channels. The modification of the $\sigma_{NN\rightarrow N\Delta}$ cross sections then takes the following form,
\begin{equation}
    \sigma^{*}_{NN\rightarrow N\Delta}(\rho) = \sigma_{NN\rightarrow N\Delta}~{e^{-\alpha^\pm (\rho/\rho_0)^{3/2}}} ,
    \label{Eq:sinel_im}
\end{equation}
where $\alpha^+$ corresponds to the constant factor to be used for $\Delta^+$ and $\Delta^{++}$ production and $\alpha^-$ is to be used for the $\Delta^0$ and $\Delta^{-}$ channels. For the in-medium $\Delta$ absorption cross sections $\sigma_{N\Delta\rightarrow NN}^{*}$, they can be determined from  the $\Delta$ production cross sections $\sigma_{NN\to N\Delta}^{*}$ by the detailed balance  relations, and they are thus modified accordingly. We then determine $\alpha^\pm$ by a fit to experimental charged pion multiplicities. This decoupling of the two classes of delta production has the advantage of also allowing for more flexibility in the subsequent decays to $\pi^-$ and $\pi^+$, leading to a more robust fit to the experimental data without any major modification to the existing dynamics.

Upon performing the fitting procedure to the charged pion multiplicities reported in {Ref.}~\cite{hades2020} for Au+Au collisions at $E_{\rm beam}=1.23$~A~GeV or $\sqrt{s_{NN}}=2.4$ GeV, values of $\alpha^+ = 0.39$ and $\alpha^-=0.70$ were found. The smaller $\alpha^+$ value found in the fit serves to reduce less the $\pi^+$ production when compared to $\pi^-$.

\section{Results}
\label{sec:results}

With the modified in-medium cross sections, $\sigma^*_{NN\rightarrow N\Delta}$, we now examine the predictions from the altered RVUU model. 

\subsection{Protons}

\begin{figure}[h]
{\includegraphics[width=\linewidth]{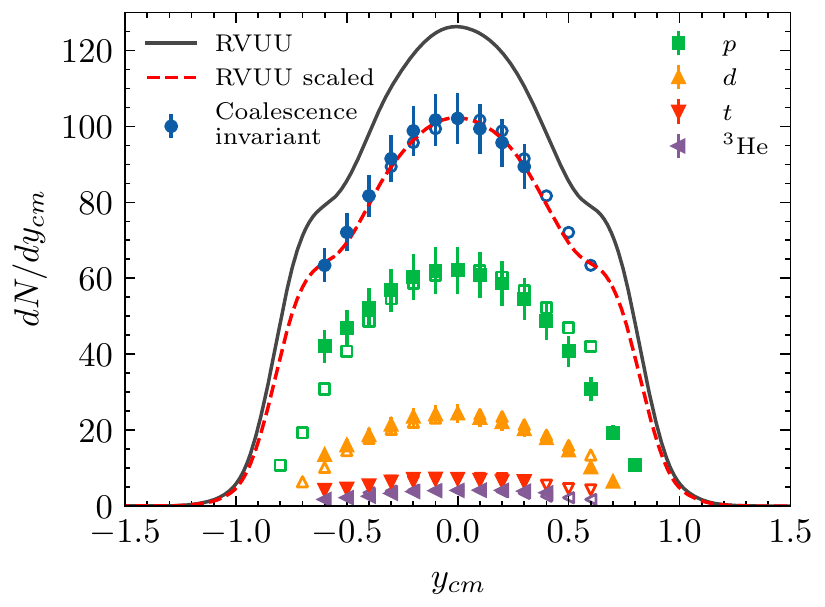}}
\caption{\label{fig:protrap}\protect(Color Online). Proton rapidity distributions from RVUU calculations (solid line) {in the center-of-mass frame of} the most central ($0\%-10\%$) {Au+Au} collisions {at $\sqrt{s_{NN}}=2.4$ GeV}. Solid squares, up-triangles, down-triangles, and left-triangles are the HADES preliminary data~\cite{Szala2019} on proton, deuteron, triton and $^3$He, respectively, and solid circles are the corresponding coalescence invariant proton rapidity distribution. Open symbols are {the reflection of} the data {with respect to the center-of-mass rapidity} $y_{cm}=0$.  The dashed line is the RVUU results multiplied by the ratio of $\frac{dN}{dy_{cm}}\vert_{y_{cm}=0}$ from the RVUU calculation and the measured coalescence invariant proton distribution. See texts for details. }
\end{figure}

Figure~\ref{fig:protrap} shows via the solid line the rapidity distribution of protons from RVUU calculations {in the center-of-mass frame of the most}  central ($0$-$10\%$ centrality) {Au+Au} collisions at $\sqrt{s_{NN}}=2.4$ GeV. Since light nuclei production is not considered in the RVUU model, to compare the proton rapidity distribution from this model with the  HADES data requires the inclusion of protons in the measured nuclei.  This so-called coalescence invariant proton rapidity distribution is given by the sum of measured rapidity distributions of all charged baryonic particles weighted with their respective charges. From the preliminary HADES data on proton, deuteron, triton and $^3$He~\cite{Szala2019}, which are shown as solid squares, up-triangles, down-triangles, and left-triangles, respectively, the resulting coalescence invariant proton rapidity distribution is shown by solid circles in Fig.~\ref{fig:protrap}.  Since the rapidity distributions of protons and these light nuclei in the center-of-mass frame of Au+Au collisions should be symmetric with respect to $y_{cm}$, we thus also show by open symbols their reflected rapidity distributions.  It is seen that the RVUU proton number is larger than the measured coalescence invariant proton number. In particular, the $dN/dy_{cm}$ at $y_{cm}=0$ from RVUU calculations and the coalescence invariant proton spectra from HADES data are about 126 and 102, respectively.  Since the present HADES data on (bound) proton rapidity distributions are still preliminary, we leave {the understanding of the about $20\%$ difference in the predicted and measured total proton numbers} for future study.  To facilitate the comparison, we also show {in Fig.~\ref{fig:protrap}} by the dashed line the RVUU results multiplied by a factor of $0.81\approx 102/126$.  The scaled RVUU results are seen to agree well with the measured coalescence invariant proton rapidity distribution.

\subsection{Pions}

\begin{table}[ht]
    \centering
    \caption{Predicted multiplicities for the $0\%-40\%$ centrality bin from the RVUU model using $\sigma_{NN\rightarrow N\Delta}$ ($\alpha^{+}=\alpha^{-}=0$)  
and $\sigma^*_{NN\rightarrow N\Delta}$ with $\alpha^{+}=\alpha^{-}=0.6$ and with $\alpha^{+}=0.39$ and $\alpha^{-}=0.7$ (see Eq.~(\ref{Eq:sinel_im})). Experimental results are from Ref.~\cite{hades2020}. }
    \label{tab:pimult}
    \begin{tabular}{c  c  c}
        \hline\hline
                    & M($\pi^-$) & M($\pi^+$) \\
                    \hline
        HADES  & $11.1\pm0.6\pm0.6$ & $6.0\pm0.3\pm0.3$ \\
        $\alpha^{+}=\alpha^{-}=0$  & 17.2 &8.7\\
        $\alpha^{+}=\alpha^{-}=0.6$ & $11.8$ & $5.4$ \\ 
        $\alpha^{+}=0.39,~\alpha^{-}=0.7$  & $11.3$ & $6.2$ \\
        \hline\hline
    \end{tabular}

\end{table}

Table~\ref{tab:pimult} shows the comparison between the pion multiplicities from the RVUU over the full $0\%-40\%$ centrality window to those reported by the HADES Collaboration in Ref.~\cite{hades2020}. The RVUU results are presented for three cases, i.e., $\alpha^+=\alpha^-=0$, $\alpha^+=\alpha^-=0.6$, and $\alpha^+=0.39$ and $\alpha^-=0.7$.  Without the in-medium reduction factor, i.e., $\alpha^+=\alpha^-=0$, the RVUU overpredicts the charged pion multiplicities by about a factor of 1.5, which is similar to the results from the other five transport models reported in Ref.~\cite{hades2020}. Using the isospin-independent in-medium reduction factors with $\alpha^+=\alpha^-=0.6$, the RVUU model overpredicts (underpredicts) the $\pi^-$ ($\pi^+$) yield, although it can reproduce the total charged pion number. Further taking into account the isospin-dependence of the in-medium reduction factors by using $\alpha^+=0.39$ and $\alpha^-=0.7$, the total number of both species of charged pions produced in the RVUU model compares well with the experimental values over the full range, indicating reasonable bulk pion production given the smaller suppression in the $\Delta^+$ and $\Delta^{++}$ channels than $\Delta^-$ and $\Delta^0$ channels. Therefore, in the following comparison with the HADES data, we will present only the RVUU results with {isospin-dependent} in-medium reduction factors of $\alpha^+=0.39$ and $\alpha^-=0.7$.

\begin{figure}[h]
\includegraphics[width=8cm]{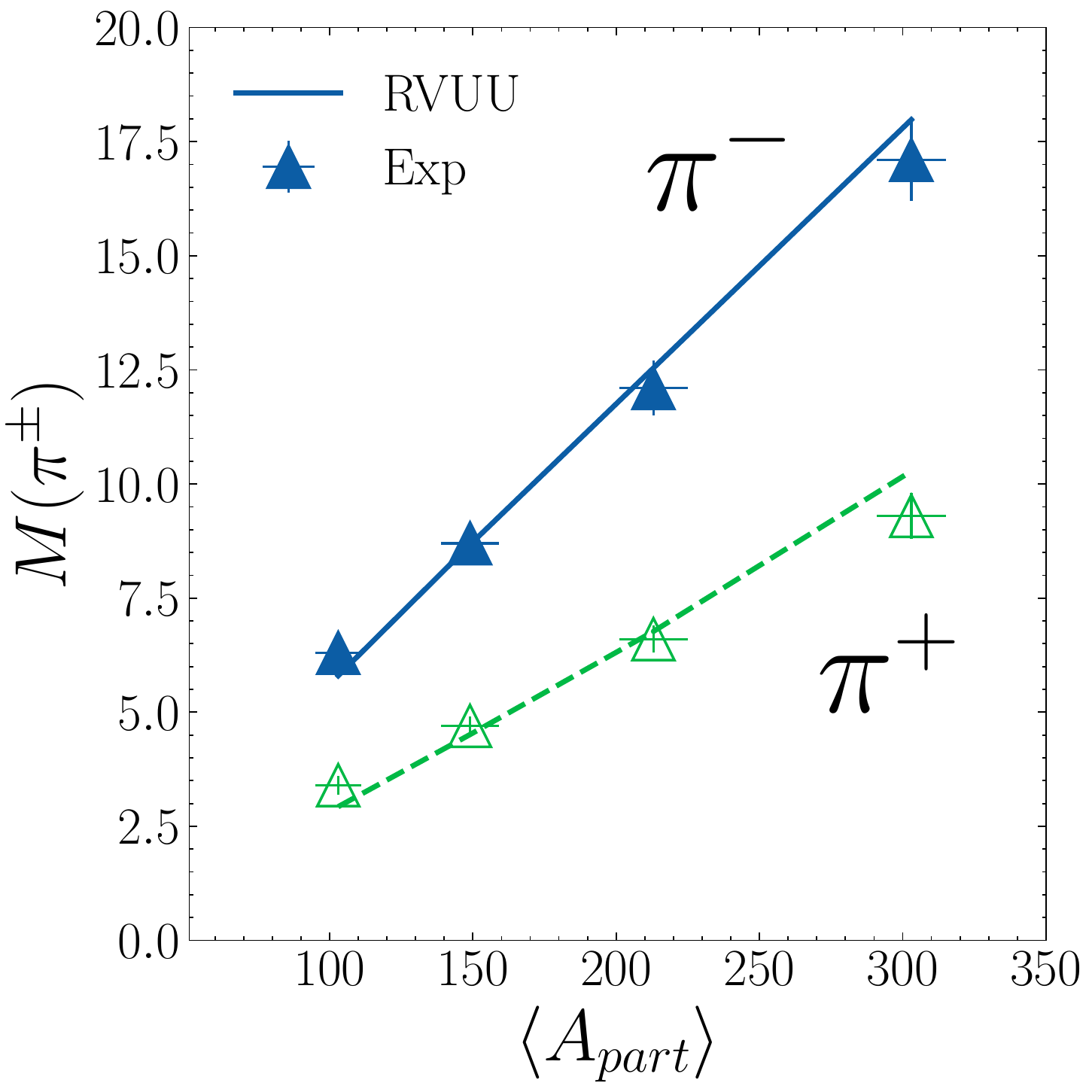}
\caption{\label{fig:apart}\protect(Color Online). Charged pion multiplicities {from} RVUU (lines) and experimental data (triangles) from {Ref.}~\cite{hades2020}. Results are shown as a function of the mean number of participants, defined in {Ref.}~\cite{hades2018}. Solid lines and filled symbols (dashed lines and empty symbols) correspond to $\pi^-$ ($\pi^+$).}
\end{figure}

To examine the scaling with the mean number of participants, $\langle A_{part} \rangle$, we plot in Fig~\ref{fig:apart} the $\pi^\pm$multiplicities within four centrality bins of $0\%-10\%$, $10\%-20\%$, $20\%-30\%$ and $30\%-40\%$, corresponding to the impact parameter ($b$) ranges of (0.0-4.7) fm, (4.7-6.6) fm, (6.6-8.1) fm, and (8.1-9.3) fm, respectively. In practice, the centrality determination and subsequent mapping of impact parameter ranges to $\langle A_{part} \rangle$ is performed in accordance with the values presented in Ref.~\cite{hades2018}.  The modification to pion production applies evenly across the range of centralities, with the behavior being roughly linear within the available range.  In comparison to the experimental data, the model predictions follow closely for all the centrality bins, except for the slight overpredictions at the most central collisions.

\begin{figure}[h]
{\includegraphics[width=4cm]{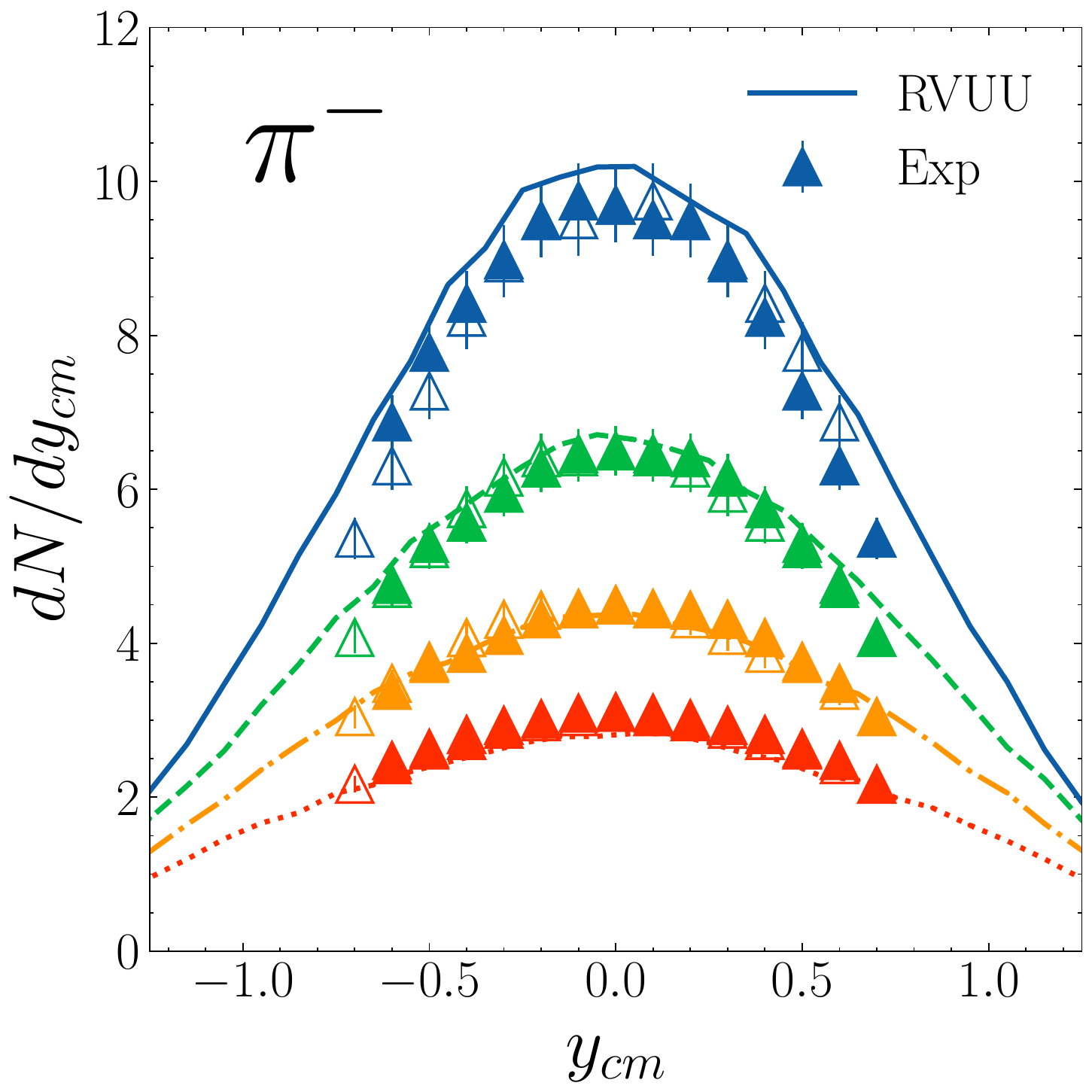}
~~\includegraphics[width=4cm]{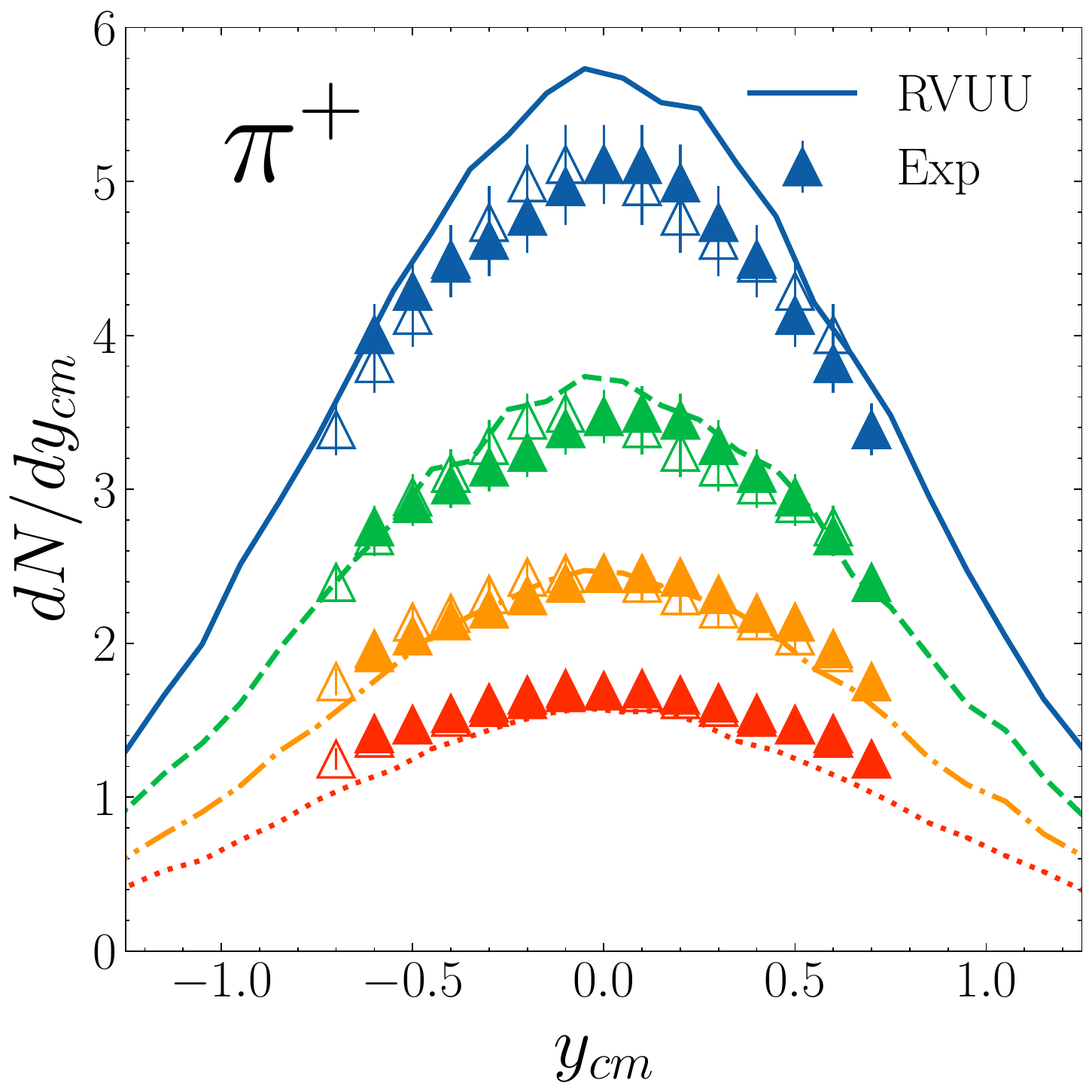}}
\caption{\label{fig:pimrap}\protect(Color Online). \protect Rapidity distributions of $\pi^-$ (left window) and $\pi^+$ (right window) from RVUU (lines) and experimental data (triangles) from {Ref.}~\cite{hades2020}. Results are shown across 4 centrality bins: $0\%-10\%$ (blue, solid line), $10\%-20\%$ (green, dashed line), $20\%-30\%$ (orange, dot-dashed line), $30\%-40\%$ (red, dotted line).}
\end{figure}

The left window of Fig.~\ref{fig:pimrap} shows rapidity distributions for negatively charged pions predicted by the RVUU model and the experimental data over the four centrality bins reported in Refs.~\cite{hades2018,hades2020}. In this case, the model reproduces the experimental data well across all centralities despite being fitted to the total $\pi^-$ multiplicity from all these classes of collisions. The results for $\pi^+$, however, exhibit significantly more variance as shown in the right window of Fig.~\ref{fig:pimrap}. Here it is seen that pions are considerably overproduced at mid-rapidity in the most central class. This behavior reflects the small variance in $\pi^+$ multiplicity presented in Fig.~\ref{fig:apart} and is unlikely to be resolved by modifications to the in-medium cross sections alone. Indeed, the fitting procedure described in Sec.~\ref{sec:methods} could be performed anew for the peripheral or central collisions separately, though this would do nothing but shift the trend the other direction. Regardless, despite the enhancement in total $\pi^+$ multiplicities or the most central ($0-10\%$) class, RVUU reproduces the approximate shape of the distribution as in the $\pi^-$ case, indicating a reasonable description of $\pi^+$ dynamics.

\begin{figure}[h]
\includegraphics[width=8cm]{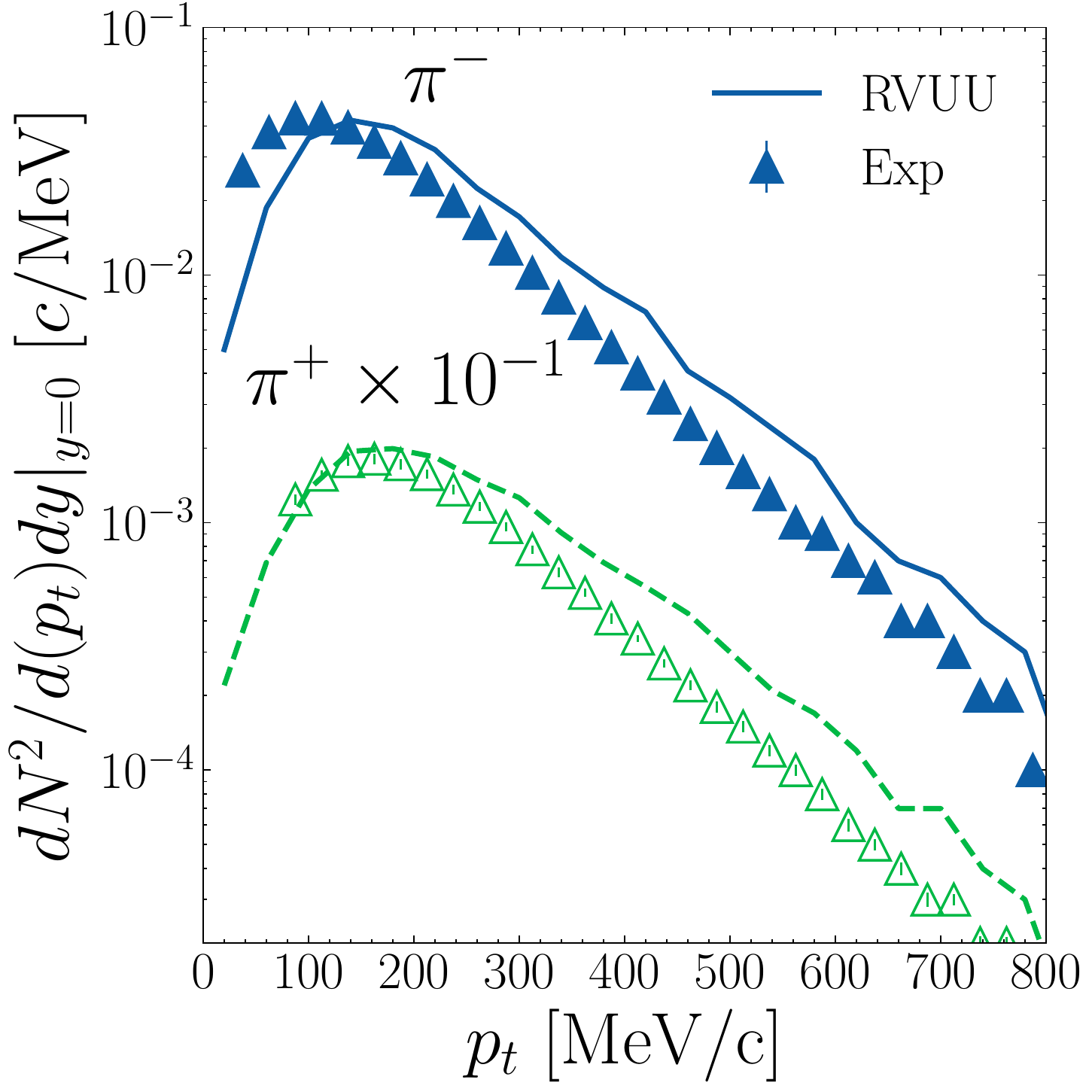}
\caption{\label{fig:pipt}\protect(Color Online). Charged pion transverse momentum spectrum from RVUU (lines) and experimental data (triangles) from {Ref.}~\cite{hades2020}. Negative pions are represented by filled blue symbols and solid blue line, while positive pions are scaled down by $10^{-1}$ and are drawn with hollow green symbols and green dashed line. Results are shown for mid-rapidity events for the most central ($0\%-10\%$) class of collisions.}
\end{figure}

Turning now to the transverse momentum $p_t$ spectrum of $\pi^\pm$, Fig.~\ref{fig:pipt} presents the model predictions along with the experimental data. It is seen that, at high $p_t$ region, the RVUU overpredicts both $\pi^-$ and $\pi^+$, and the slopes from RVUU calculations are more gradual than observed in the data.  At low $p_t$ region, the  $\pi^-$ is underpredicted by RVUU.  This deviation could be due to the absence of the pion mean-field potential, which enhances the production of pions at low $p_t$ ~\cite{Estee:2021khi} and further affects the fit of the in-medium $\Delta$ production cross section.

\section{Conclusions}
\label{sec:conclusions}

In the present study, we have used the isospin-dependent RVUU model to study the production of charged pions from Au+Au collisions at $\sqrt{s_{NN}}=2.4$ GeV. With the medium dependence of the $\Delta$ resonance production cross section from the nucleon-nucleon inelastic scattering determined by fitting the total multiplicities of $\pi^-$ and $\pi^+$ measured in the HADES experiment, we have obtained a good description of the rapidity distributions of both $\pi^-$ and $\pi^+$ for various centrality bins. For the transverse momentum spectra, the RVUU underpredicts $\pi^-$ at low $p_t$, while overpredicts $\pi^+$ and $\pi^-$ at high $p_t$ region.  We have attributed this discrepancy to the absence of the pion mean-field potential in the RVUU model. The reasonable success of the RVUU model in describing the HADES pion data is in contrast to the results from other transport models~\cite{Hartnack:1997ez,Cassing:1999es,Aichelin:2019tnk,Buss:2011mx,Petersen:2018jag} used by the HADES Collaboration to compare with its data, which all overestimate the $\pi^-$ and $\pi^+$ multiplicities for all centralities by factors ranging from 1.2 to 2.1.  The source of this difference between our results and those from other transport models are mainly due to our introduction of a density-dependent reduction factor to the nucleon-nucleon inelastic cross section, which is absence in other models. However, the introduced in-medium reduction factors for $\Delta^+$ and $\Delta^0$ production lead to a considerably larger splitting of their cross sections compared with theoretical calculations in Refs.~\cite{Cui2021,LI2017557}. To pin down the in-medium $NN$ inelastic cross sections requires further theoretical studies.  

\begin{acknowledgments}
The authors  acknowledge helpful discussions with Yongjia Wang. This work was supported in part by the U.S. Department of Energy under Award Nos. DE-SC0015266 (C.M.K.) and DE-SC0013365 (K.G.), the Welch Foundation under Grant No. A-1358 (K.G. and C.M.K.), the National Natural Science Foundation of China under Grant No. 11905302 (Z.Z.), Guangdong Basic and Applied Basic Research Foundation under Grant No. 2019A1515010849 (Z.Z.), and the National Science Foundation under Grant No. PHY1652199 (J.W.H.). K.G. acknowledges partial support through computational resources and services provided by the Institute for Cyber-Enabled Research at Michigan State University.
\end{acknowledgments}

\bibliography{PionIME0429}

\end{document}